# A New Concept for Direct Imaging and Spectral Characterization of Exoplanets in Multi-planet Systems

Taro Matsuo*[a], Wesley A. Traub[b], Makoto Hattori[c], and Motohide Tamura[a]
[a]National Astronomical Observatory of Japan, 2-21-1, Osawa, Mitaka, Tokyo, Japan 181-0015;
[b]Jet Propulsion Laboratory, California Institute of Technology 4800 Oak Grove Drive, Pasadena, CA 91109, USA;
[c] Tohoku University, 6-3 Aoba, Aramaki, Aoba-ku, Sendai 980-8578;

**ABSTRACT**

We present a novel method for direct detection and characterization of exoplanets from space. This method uses four collecting telescopes, combined with phase chopping and a spectrometer, with observations on only a few baselines rather than on a continuously rotated baseline. Focusing on the contiguous wavelength spectra of typical exoplanets, the (u,v) plane can be simultaneously and uniformly filled by recording the spectrally resolved signal. This concept allows us to perfectly remove speckles from reconstructed images. For a target comprising a star and multiple planets, observations on three baselines are sufficient to extract the position and spectrum of each planet. Our simulations show that this new method allows us to detect an analog Earth around a Sun-like star at 10pc and to acquire its spectrum over the wavelength range from 8 to 19μm with a high spectral resolution of 100. This method allows us to fully characterize an analog Earth and to similarly characterize each planet in multi-planet systems.

## 1. INTRODUCTION

Since the first discovery of an exoplanet orbiting a main-sequence star by Mayor & Queloz (1995), more than 500 exoplanets have been discovered so far through radial velocity and several other ongoing indirect observations. On the other hand, the first success of direct imaging was reported 12 years after the first detection of an exoplanet. Marois et al. (2008) and Kalas et al. (2008) imaged exoplanets orbiting A-type Vega-like stars of HR8799 and Fomalhaut with Keck/Gemini and Hubble Space Telescope, respectively. Following these successes, a snapshot of a planet candidate orbiting a fairly old G star, GJ758B, has been taken with Subaru/HiCIAO (Thalman et al. 2009). In addition, Lagrange et al. (2010) confirmed a planet around beta Pic with Very Large Telescope, originally reported in 2008. Thus, the number of directly imaged exoplanets has recently increased but extraordinary efforts are required for direct detection of even these young self-luminous gas giants. However, direct detection of the Earth-like planets in the habitable zones of other planetary systems involves a more significant technical challenge, compared with the direct imaging of young gas giants. This is because much higher contrast values of $10^{-10}$ and $10^{-7}$ at a close angular distance to a central star are required at visible and infrared wavelengths, respectively. Several possibilities for direct detection of the Earth-like planets have been proposed: space-based coronagraphs operating at visible wavelengths, space-based nulling interferometers operating at mid-infrared wavelengths. In this paper, we concentrate on the latter approach.

Bracewell (1978) proposed the simplest nulling interferometer for direct imaging of Earth-like planets in the habitable zones at mid-infrared wavelengths. This type of interferometer comprises two collecting apertures separated by baseline length $B$, and introduces an achromatic phase-shift of $\pi$ between the two arms such that the light from an on-axis source is canceled at the beam combiner output. The transmission pattern on the sky is a sinusoidal corrugation with a null at the center of the field of view and an angular periodicity of $\lambda/B$, where $\lambda$ is the observing wavelength. Regarding the null depth, Angel and Woolf (1997) proposed some improvements to the single-Bracewell configuration. However, even if the on-axis source is perfectly suppressed with the nulling technique, emissions from the local zodiacal and the exo-zodiacal dust disks are much brighter than those from Earth-like planets in the mid-infrared wavelength range.

A first step toward detecting a planet against a bright zodiacal background is the sine phase chopping technique proposed by Mennesson and Mariotti (1997). The method can separate asymmetrical sources, such as the planets, from symmetrical ones such as the central stars, local zodiacal lights, and exozodiacal light. A dual-Bracewell configuration, which is composed of two single nulling interferometers combined with the phase chopping technique, overcomes the disadvantages of the original single-Bracewell configuration. This type of interferometer can clearly identify the location of a planet through the maximum correlation method, analogous to Fourier transformation used for standard radio interferometers. However, the relative phase between successive rolls needs to be maintained. Otherwise, speckles are produced in a reconstructed image.

Previously, Matsuo et al. (2010) proposed a new concept for direct detection and spectral characterization of the Earth-like planets by focusing on the fact that observing wavelengths can be contiguously obtained in space. Thus, the (u, v) plane can be simultaneously filled by observing spectra at only a few discrete orientations instead of continuously rotating the baselines, which is the case with the original Bracewell approach. There are several advantages to this concept. First, observations on two baselines are sufficient to extract the position of a single planet and its spectrum. Second, the higher spectral resolution leads to more dense filling of the (u,v) plane. Third, the phase at each roll angle is irrelevant in this method. As a result, there is no speckle in the reconstructed image. Thus, this concept could be a promising method for direct detection of exoplanets. However, it cannot be directly applied to multi-planet systems, such as Gliese 876 (e.g., Rivera et al. 2010), 55Cancri (e.g., McArthur et al. 2004), and Gliese 581 (e.g., Mayor et al. 2009), as explained next.

In this paper, we extend the original concept proposed by Matsuo et al. (2010) for the multi-planet systems. For these, observations on two baselines are not sufficient to estimate each planet's position even if the separations of each signal from the host star along orthogonal two axes can be clearly extracted from two one-dimensional images of the planetary system. This is because there is an ambiguity in extracting the two-dimensional positions of more planets, given only two projected images on orthogonal axes. In Section 3 we perform its numerical simulation to validate the concept. In Section 4 we show how to extract exoplanet spectra and discuss the capability of characterization of the Earth-like planets, and in Section 5 we give our conclusions.

## 2. THEORY

In this section, we expand the original concept proposed by Matsuo et al. (2010) to include the multi-planet systems. The method for extraction of each planet's position is presented in Section 2.1. The spectrum of each planet is analytically derived in Section 2.2. An improvement of signal-to-noise ratio (S/N) for the spectrum of each planet is introduced in Section 2.3. Finally, the field of view of this method is discussed. We assume that the target is composed of a star and $n$ multiple planets. A dual-Bracewell nulling interferometer with a sine phase chopping and a spectrometer is assumed. We assume that the exozodiacal light has a symmetric brightness distribution with respect to the star.

### 2.1 Direct Imaging of a Planetary System

Based on the previous studies (e.g., Lay 2004; Draper et al. 2006), the spectrally resolved signal, in unit of detected photoelectrons of the dual-Bracewell nulling interferometer for the individual chop state, is expressed as

$$O_{\pm}(\lambda) = \frac{1}{2} \iint d^2\theta I(\lambda,\vec{\theta}) \sin^2\left(\frac{\pi}{\lambda}\vec{b}\cdot\vec{\theta}\right)\left\{1 \pm \sin\left(\frac{2\pi}{\lambda}\vec{B}\cdot\vec{\theta}\right)\right\} \qquad (1)$$

where $\vec{\theta}$ is the position vector on the sky, $I(\lambda,\vec{\theta})$ is signal from the source on the sky at $\lambda$ with a single-dish telescope, $\vec{b}$ is the nulling baseline vector, $\vec{B}$ is the imaging baseline vector. The $I(\lambda,\vec{\theta})$ represents a product of the intensity of the source, total correcting area, bandwidth in a spectral element, detector efficiency, and the beam efficiency. The + and - notations express the plus (+) and minus (-) chop states, respectively.

Here, in order to apply this concept to multip-planet systems, we assume that a target comprises a star, $n$ planets, and local zodiacal plus exozodiacal light. The spectrally resolved signal is written as

$$O_{\pm}(\lambda) = \frac{1}{2} \iint_{\Omega_*} d^2\theta I_*(\lambda,\vec{\theta}) \sin^2\left(\frac{\pi}{\lambda}\vec{b}\cdot\vec{\theta}\right)\left\{1 \pm \sin\left(\frac{2\pi}{\lambda}\vec{B}\cdot\vec{\theta}\right)\right\}$$
$$+ \frac{1}{2}\sum_{k}^{n} I_{p,k}(\lambda,\vec{\theta}_{p,k})\Omega_{p,k} \sin^2\left(\frac{\pi}{\lambda}\vec{b}\cdot\vec{\theta}_{p,k}\right)\left\{1 \pm \sin\left(\frac{2\pi}{\lambda}\vec{B}\cdot\vec{\theta}_{p,k}\right)\right\} \qquad (2)$$
$$+ \frac{1}{2} \iint_{\Omega_{fov}} d^2\theta\left\{I_{local-zodi}(\lambda) + I_{exozodi}(\lambda,\vec{\theta})\right\} \sin^2\left(\frac{\pi}{\lambda}\vec{b}\cdot\vec{\theta}\right)\left\{1 \pm \sin\left(\frac{2\pi}{\lambda}\vec{B}\cdot\vec{\theta}\right)\right\}$$

where $\vec{\theta}_{p,k}$ is the position vector of the $k$th planet. $\Omega_*$ and $\Omega_{p,k}$ are the solid angles subtended by the star and the $k$th planet, and $\Omega_{fov}$ is the field of view of this nulling interferometer. $I_*(\lambda,\vec{\theta})$, $I_{p,k}(\lambda,\vec{\theta}_{p,k})$, $I_{local-zodi}(\lambda)$, and $I_{exozodi}(\lambda,\vec{\theta})$ are the intensities of the star, the $k$th planet, the local zodiacal light, and the exozodiacal light at $\lambda$, respectively. The demodulated signal is

$$O(\lambda) = O_+(\lambda) - O_-(\lambda)$$
$$= \iint_{\Omega_*} d^2\theta I_*(\lambda,\vec{\theta}) \sin^2\left(\frac{\pi}{\lambda}\vec{b}\cdot\vec{\theta}\right)\sin\left(\frac{2\pi}{\lambda}\vec{B}\cdot\vec{\theta}\right)$$
$$+ \sum_{k}^{n} I_{p,k}(\lambda,\vec{\theta}_{p,k})\Omega_{p,k} \sin^2\left(\frac{\pi}{\lambda}\vec{b}\cdot\vec{\theta}_{p,k}\right)\sin\left(\frac{2\pi}{\lambda}\vec{B}\cdot\vec{\theta}_{p,k}\right) \qquad (3)$$
$$+ \iint_{\Omega_{fov}} d^2\theta\left\{I_{local-zodi}(\lambda) + I_{exozodi}(\lambda,\vec{\theta})\right\} \sin^2\left(\frac{\pi}{\lambda}\vec{b}\cdot\vec{\theta}\right)\sin\left(\frac{2\pi}{\lambda}\vec{B}\cdot\vec{\theta}\right)$$

Since the stellar leakage and the local as well as exozodiacal light are centro-symmetric, the above equation reduces to

$$O(\lambda) = \sum_{k}^{n} I_{p,k}(\lambda,\vec{\theta}_{p,k})\Omega_{p,k} \sin^2\left(\frac{\pi}{\lambda}\vec{b}\cdot\vec{\theta}_{p,k}\right)\sin\left(\frac{2\pi}{\lambda}\vec{B}\cdot\vec{\theta}_{p,k}\right). \qquad (4)$$

We now show that the (u, v) plane can be simultaneously and uniformly filled by the use of the contiguous observing wavelengths. Fourier transformation of the spectrally resolved signals along the observing wavelengths gives a one-dimensional image of the planetary system:

$$M(\vec{\alpha}) = \int_{\lambda_{min}}^{\lambda_{max}} d\lambda O(\lambda) \sin\left(\frac{2\pi}{\lambda}\vec{B}\cdot\vec{\alpha}\right), \qquad (5)$$

where $\vec{\alpha}$ is position vector on the sky, and $\lambda_{min}$ and $\lambda_{max}$ are the minimum and maximum observing wavelengths, respectively. On a related issue, Ohta et al. (2006) established a theory for imaging of extended sources in radio, focusing on broadband ground-based radio observations. From the reconstructed image of the planetary system, we can estimate the separation of each signal from the host star along the imaging baseline vector. As described in Matsuo et al. (2010), if the target is a single planet system, the two-dimensional position of the planet could be estimated from the two one-dimensional images of the planetary system. However, the observations on two baselines are not sufficient for extraction of the position of each planet because no one knows which signal appearing in one of the two reconstructed images corresponds to which signal in the other reconstructed image. As a result, the separations of each signal from the host star along the two orthogonal axes cannot be transformed in principle into the position of each signal in the two-dimensional plane.

We can break this degeneracy with two planets by adding a third position of the array. The coordinate system on the sky is set as a rectangular Cartesian coordinate system, (x, y). We assume that the separation of each planet from the host star along the two orthogonal axes is already known from the first and second observations, and the rotation angles of the array in the third measurement and the $k$th planet are $\phi_B$ and $\phi_{p,k}$ from the $x$-axis, respectively; the pair of the separations of the $k$th planet from the host star along the two orthogonal axes is satisfied by

$$\frac{\overline{a}_{p,k}}{\cos(\phi_B - \phi_{p,k})} = \sqrt{x_{p,k}^2 + y_{p,k}^2}, \tag{6}$$

where $\overline{a}_{p,k}$ is the measured separation of the $k$th planet from the host star along the imaging baseline in the third measurement, $x_{p,k}$ and $y_{p,k}$ are the separation of the $k$th planet along the $x$- and the $y$-axes, respectively, and $\phi_{p,k}$ is

$$\phi_{p,k} = \tan^{-1}\left(\frac{x_{p,k}}{y_{p,k}}\right). \tag{7}$$

Here, no one knows the right pair of separations of each planet from the host star along the two orthogonal axes. Therefore, we apply all of the combinations of the pairs to Equation (6) and then choose a combination of them which is best satisfied by Equation (6). The chosen pairs represent the separations of planets from the host star along the two orthogonal axes. Thus, observations on three baselines are sufficient for extraction of the two-dimensional position of each planet. On the other hand, the information on the spectrum of each planet is lost through Fourier transformation of the spectrally resolved signal along the observing wavelength. Therefore, the spectrum of the planet should be reconstructed for characterization of the planet, as we discuss next.

## 2.2 Spectrum of Each Planet

In the section, we present a new method for reconstruction of the spectrum of each planet from its two-dimensional position of each planet from its two-dimensional position. Given that the number of the planets within the field of view is $n$ and the two-dimensional position of each planet is clearly revealed through the procedure shown in the previous section, the demodulated spectrally resolved signal at $\lambda$ can be rewritten as

$$O(\lambda) = O_*(\lambda) + O_{local-zodi}(\lambda) + O_{exozodi}(\lambda) + \sum_k^n O_{p,k}(\lambda), \tag{8}$$

where $O_*(\lambda)$, $O_{local-zodi}(\lambda)$, $O_{exozodi}(\lambda)$, and $O_{p,k}(\lambda)$ are the demodulated spectrally resolved signals of the stellar leakage, the local zodiacal light, the exozodiacal light, and the $k$th planet at $\lambda$, respectively. Here, as described in the previous section, the star, the local zodiacal light, and the exozodiacal light are centro-symmetric sources. As a result, we can express Equation (8) as

$$O(\lambda) = O_s(\lambda) + \sum_k^n O_{p,k}(\lambda), \tag{9}$$

where $O_s(\lambda)$ represents the sum of the demodulated signals from the symmetric sources. Keeping the array configuration unchanged during rotation of the array on the optical axis of the array, the signal of the symmetric source is unchanged:

$$O_{s,1}(\lambda) \approx O_{s,2}(\lambda), \tag{10}$$

where $O_{s,n}(\lambda)$ is the demodulated signal of the symmetric source in the $n$th measurement. On the other hand, the planet as an off-axis source produces the modulated signal during rotation of the array. Therefore, the observations on the rotation of the array can separate the planets from the centro-symmetric sources. The observations on $n+1$ baselines are required for extraction of the spectra of $n$ planets:

$$\begin{pmatrix} O_1(\lambda) \\ \vdots \\ O_{n+1}(\lambda) \end{pmatrix} = Q \begin{pmatrix} O_s(\lambda) \\ \vdots \\ I_{p,n}(\lambda) \end{pmatrix}, \tag{11}$$

and then

$$\begin{pmatrix} O_s(\lambda) \\ \vdots \\ I_{p,n}(\lambda) \end{pmatrix} = Q^{-1} \begin{pmatrix} O_1(\lambda) \\ \vdots \\ O_{n+1}(\lambda) \end{pmatrix}. \tag{12}$$

The element of the matrix $Q$ in the $l$th row and the $m$th column, except for the first column, is expressed as

$$q_{l,m} = \sin^2\left(\frac{\pi}{\lambda}\vec{b}_l \cdot \vec{\theta}_m\right)\sin\left(\frac{2\pi}{\lambda}\vec{B}_l \cdot \vec{\theta}_m\right), \tag{13}$$

where $\vec{B}_l$ and $\vec{b}_l$ are respectively the imaging and the null baseline vectors in the $l$th measurement, and the $\vec{\theta}_m$ is the position angle vector of the $m$th planet. On the other hand, the elements of the matrix $Q$ in the first column are 1. The reconstructed spectrum of the $m$th planet is

$$I_{p,m}(\lambda) = \sum_{k}^{n} q'_{m,k} O_k(\lambda), \tag{14}$$

where $q'_{l,m}$ represents the element of the inverse matrix $Q^{-1}$ in the $l$th row and the $m$th column. When $n=1$, the above equation can be rewritten as

$$I_{p,1}(\lambda) = \frac{O_2(\lambda) - O_1(\lambda)}{\left\{\sin^2\left(\frac{\pi}{\lambda}\vec{b}_2 \cdot \vec{\theta}_{p,1}\right)\sin\left(\frac{2\pi}{\lambda}\vec{B}_2 \cdot \vec{\theta}_{p,1}\right) - \sin^2\left(\frac{\pi}{\lambda}\vec{b}_1 \cdot \vec{\theta}_{p,1}\right)\sin\left(\frac{2\pi}{\lambda}\vec{B}_1 \cdot \vec{\theta}_{p,1}\right)\right\}}. \tag{15}$$

This result is equal to that derived by Matsuo et al. (2010). On the other hand, the sky transmission of each planet in the multi-planet system is very complicated but it can be derived analytically. In general, the *SNR* is

$$SNR \approx \frac{I_p(\lambda)}{2\sqrt{O(\lambda)}}, \tag{16}$$

where $\sqrt{O(\lambda)}$ is the photon variance of the total signal per spectral element in each chop state. The factor of 1/2 in Equation (16) results from subtracting the (+) and (-) chop states. Here, the wavelength dependence of the sky transmission for each planet is determined from the position of each planet and the imaging and the null baseline vectors. Generally, the dependency is much stronger as the number of the planets increases.

### 2.3 Observations on $n+2$ baselines

We introduce an additional observation to avoid the wavelength dependency of the sky transmission because the number of independent acquired spectra corresponds to the combination of the observations. For simplicity, when $n = 1$, we take observations on three baselines and independently acquire the planet's spectrum three times. Here, these spectra are weighted with *SNR* and then are summed:

$$I(\lambda)_{average} = \frac{\{SNR_{1-2}(\lambda) \times I_{1-2}(\lambda)\} + \{SNR_{2-3}(\lambda) \times I_{2-3}(\lambda)\} + \{SNR_{3-1}(\lambda) \times I_{3-1}(\lambda)\}}{SNR_{1-2}(\lambda) + SNR_{2-3}(\lambda) + SNR_{3-1}(\lambda)}, \tag{17}$$

where $SNR_{m-n}(\lambda)$ and $I_{m-n}(\lambda)$ are the *SNR* and the planet's spectrum derived by combination of the $m$th and $n$th observations. The averaged *SNR* is defined as

$$SNR_{average}(\lambda) = \sqrt{(SNR_{1-2})^2 + (SNR_{2-3})^2 + (SNR_{3-1})^2}. \tag{18}$$

As shown in Equation (18), the wavelength dependency of *SNR* could be relaxed, and then the *SNR* would improve over the whole observed wavelength range. Thus, observations on *n*+2 baselines can more effectively characterize *n* exoplanets. In this case the matrix *Q* is not square, and the inverse is replaced by the singular value decomposition (SVD) matrix.

## 2.4 Field of View

Finally, we discuss the field of view of this method. Generally, the field of view of stellar interferometers are characterized by the sampling theorem. In other words, the field-of-view is limited by an incomplete coverage of the (u, v) plane. On the other hand, according to this new method, the spectral resolution corresponds to incomplete coverage of the (u, v) plane because the image of the planetary system can be reconstructed through Fourier transformation along the observing wavelengths. As described by Matsuo et al. (2010), the radius of the field-of-view is limited by the spectral resolution:

$$\theta_{FOV} = \frac{\lambda_{ave}}{B} \frac{\lambda}{\delta\lambda}.  \quad (19)$$

where $\lambda_{ave}$ is the average of the observing wavelengths and $\frac{\lambda}{\delta\lambda}$ is the spectral resolution of this system. In the case that the spectral resolution is 100 and the imaging baseline length is 100m, the field-of-view of this system is 2 arcsec.

## 3. SIMULATION

We performed numerical simulations to validate the proposed method. Based on the method, an image of a target planetary system is extracted and then the spectrum of each planet is estimated. In this section, we introduce the parameters required for the numerical simulations and show the reconstructed image of the target system and the estimated spectrum of each planet.

### 3.1 Parameters

The target system is composed of a star, two planets, and a symmetric exozodiacal light distribution. The distance of this target system is 10pc. The target star is a Sun-like star (G2V) with an effective temperature $T_s$ = 5784K and one solar luminosity. The physical size of the target star is one solar radius. These parameters are same as those of the Sun. One of the target planets, defined as "P1", is taken to be an analog Earth with an Earth radius $R_E$ = 6300km. It is positioned at (x, y) = (0.6AU, 0.8AU) around the star, where (x, y) is the coordinate system on the sky. The orbital radius of the planet corresponds to 1.0AU. The atmospheric composition of the target planet includes ozone, carbon dioxide, water, and methane. Another target planet, defined as "P2", is assumed to be a super-Earth with a radius of two Earth radii. The position of another target planet is at (1.6AU, 1.2AU) around the star. The semi-major axis of the planet is 2.0AU. The effective temperature of the planet is determined through its thermal equilibrium: $T_P = 265K \left(\frac{a}{1AU}\right)^{-1/2} \left(\frac{L}{1L_\odot}\right)^{1/4}$, where $a$ is the semi-major axis of planet and $L$ is the luminosity of host star. The effective temperatures of the planets P1 and P2 are 265K and 187K, respectively. The spectrum of the P2 planet is taken to be a blackbody. The azimuth angles of the planet P1 and P2 from the x-axis are defined as $\varphi_{P1}$ and $\varphi_{P2}$. A diagram of the system is shown in Figure 2.

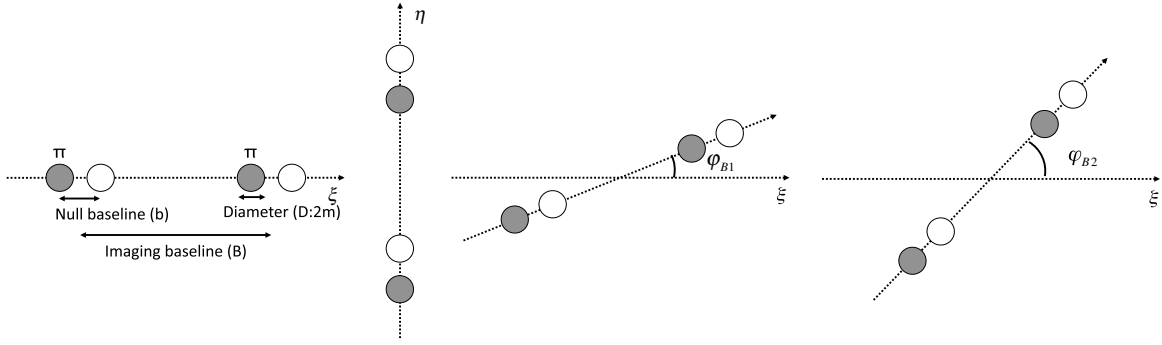

Figure 1. Array configuration for each measurement. The azimuth angles of the array in the third measurement and the forth measurement from the $\xi$ axis are set as $\varphi_{B1}$ and $\varphi_{B2}$, respectively. The $\xi$ axis is parallel to the *x*-axis on the sky.

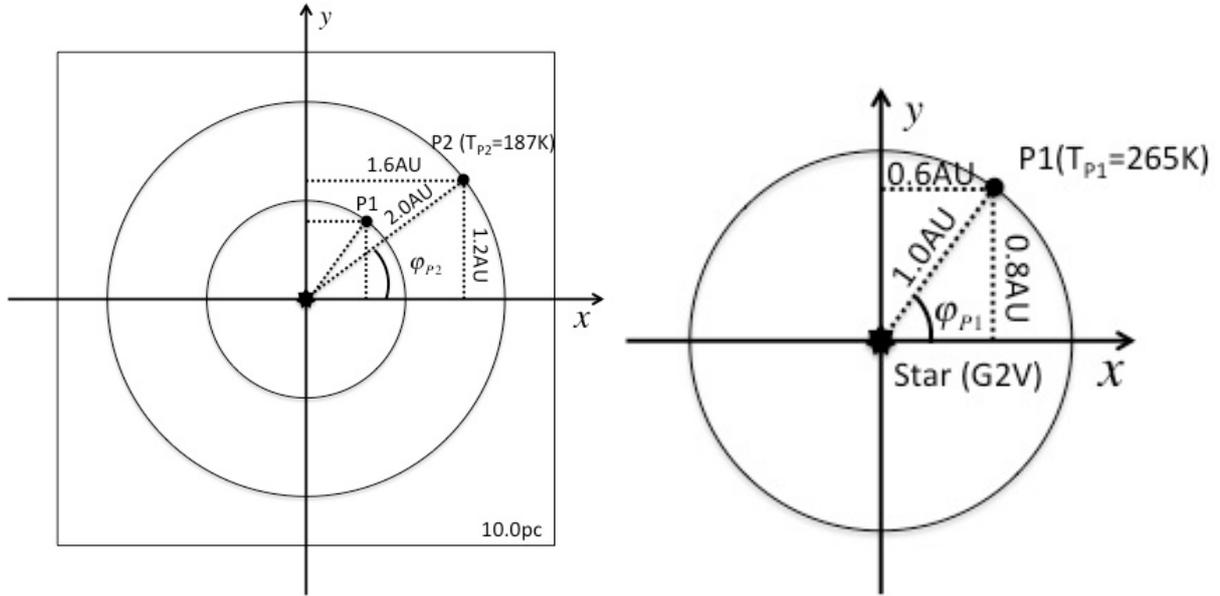

Figure 2. Conceptual diagram of the target planetary system (left) and the center region of the target system (right). P1 and the P2 represent an analog Earth and super-Earth, respectively.

In addition, we apply a dual-Bracewell nulling interferometer with four 2.0m diameter correcting mirrors to this system (see Figure 1). The coordinate system on the aperture plane is $(\xi,\eta)$, where the $\xi$- and $\eta$- axes are parallel to the x- and y- axes on the sky, respectively. The peak-to-peak pointing jitter of the array is set as 0.50mas based on a previous study (Beichman & Velusamy 1999). This pointing jitter of the array leads to breaking of the above centro-symmetric assumption. In other words, the target star can no longer be considered as an on-axis star, and an offset of the exozodiacal cloud produces an asymmetric brightness distribution. As a result, the stellar leakage becomes very large at wavelengths less than 10μm but the offset does not significantly affect the signal. We also consider the quantum efficiency, the read noise, and the dark current of the detector. Assuming that the detector array used for this simulation is the Si:As Impurity Band Conductor arrays developed for JWST MIRI, the quantum efficiency is 0.8, the read noise is 19e-, and the dark current is 0.03e-/s (Love et al. 2004). The photoelectron signals of the target star, the exozodiacal light, and the local zodiacal light in one-hour integration time are shown in Figure 3.

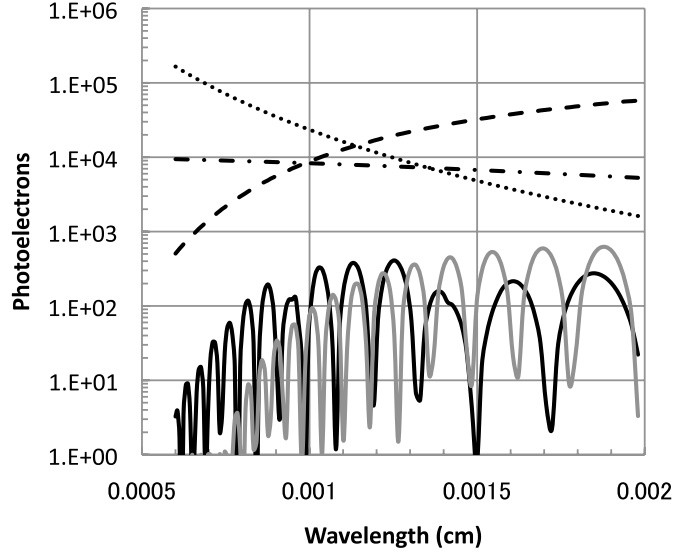

Figure 3. Signals with a spectral resolution R = 100 in one-hour integration time for one of the two chop states. The black and the gray solid lines are the spectra of the analog Earth and the super-Earth, respectively. The dotted, dash-dotted, and dashed lines represent the spectra of the stellar leakage, the exozodiacal light, and the local zodiacal light, respectively. The null depth of this system is $10^{-5}$ at 10 μm when the null and the imaging baseline lengths are 14m and 300m, respectively. The fluxes of the local zodiacal light and the exozodiacal light are estimated, based on previous studies (e.g., Beichman & Velusamy 1999). We assume that the amount of exozodiacal light is three times higher than that of our solar system. The Earth-like planet shows methane absorption at 7.4μm, ozone at 9.6μm, carbon dioxide at 15μm, and water absorptions at 6.3 μm and beyond 12 μm while the super-Earth planet is taken to be just a blackbody object.

## 3.2 Position of Each Planet

First, we perform the observations on three baselines to extract the position of each planet. Based on this method, we rotate the array by 90 degrees after the first measurement and then run the second measurement. Additionally, we rotate the array by $\varphi_{B1}$ from the $\xi$-axis to determine the positions of the two target planets on the sky. Figure 1 shows the array configuration and the azimuth angle of the array for each measurement. Here, the imaging baseline length is set as 300m for accurate measurement. The null baseline is set as 14m for suppression of the stellar leakage. The transmission of the target planet positioned on the inner limit of habitable zone is 0.5 at 8μm. The inner limit of habitable zone is assumed to be 0.6AU in the target system. For simplify, we assume that the imaging baselines of the array in the first and the second measurements are perfectly parallel to the *x*-axis and the *y*-axis on the sky, respectively (See Figure 2). The azimuth angle of the array in the third measurement, $\varphi_{B1}$, is set as $\frac{4\pi}{9}$. According to Equation (4), Fourier transformation of the spectrally resolved signals along the observing wavelengths for each measurement gives the one-dimensional position of the planet projected by each imaging baseline.

We use the signals at the wavelengths between 9 and 18μm but the observing wavelength range we assume is 6 - 20μm. This is because the stellar leakage is large at the wavelengths less than 8μm due to the pointing jitter of the array and the local zodiacal light is very large at wavelengths greater than 18μm. Figure 4 shows the reconstructed image of the planetary system through Fourier transformation of the spectrally resolved signals. An enlarged view of the central region of Figure 4 is shown in Figure 5. As shown in Figure 4 and 5, there appear two strong peaks, which are positioned at 0.6AU and 1.6AU for the first measurement, 0.8AU and 1.2AU, and 0.89AU and 1.46 AU for the third measurement. The positions of the two planets on the sky are determined such that the separations of the fringe peaks from the host star along the imaging baseline vectors are satisfied with Equation (5). As a result, the positions of the Earth-like planet and the super Earth-like planet are respectively obtained as (0.6AU, 0.8AU) and (1.6AU, 1.2AU) from our simulation. Here, the error in the position of each planet is determined by the spatial resolution, which is defined as $\lambda/2B$. In this simulation, because the FWHMs of the fringes are about 0.04 AU, the error in the position of each planet is $\pm 0.02$ AU from the observation on one baseline. Therefore, from observations on more than three baselines, the maximum error in the position of the planet for each projected semimajor axis is $0.02/\sqrt{2}$ AU. On the other hand, as shown in Figure 4, the noise increases beyond 5.0AU, for each measurement. This comes from the incomplete coverage of the (u, v) plane. In other words, the field-of-view in this case is around 1.0 arcsec, corresponding to that derived from Equation (17). In summary, we can measure the planets within the field of view and accurately determine the two-dimensional position of each one. It requires at least half a day to extract the image of the multiple planet system at 10pc.

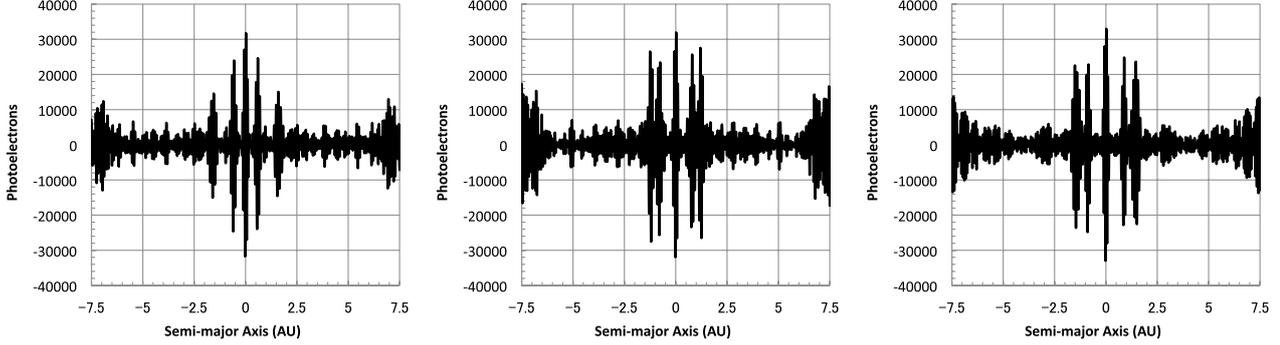

Figure 4. Images of the planetary system along the *x*-axis (left) and *y*-axis (right). The integration time for each baseline is 12 hr. The target planets are positioned at (x, y) = (0.6 AU, 0.8 AU).

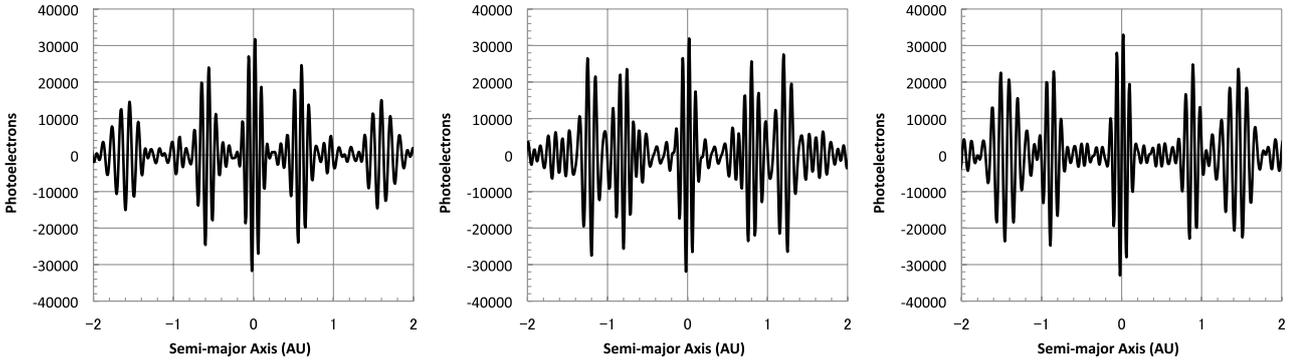

Figure 5. Enlarged views of the central region, $-2.0 AU < a < 2.0 AU$ of Figure 4.

### 3.3 Spectrum of Each Planet

Next, we reconstruct the spectrum of each planet for its characterization. In this phase, the long imaging baseline is no longer required because the position of the target planet is accurately determined. Therefore, we set the imaging baseline length as 40m for more suppression of the stellar leakage while we keep the null baseline length unchanged. As in the case of the observations for reconstruction of the image, the imaging baselines in the first and the second measurements are parallel to the *x*- and the *y*-axes, respectively. According to Equation (11), the spectrum of each planet can be reconstructed from the observations on three baselines. We set the azimuth angle of the array in the third measurement, $\varphi_{B1}$, as $\frac{5\pi}{9}$. Figure 6 shows the wavelength dependence of the *SNR* for each planet. The *SNR* of the reconstructed spectrum for each planet is more than five at the very limited wavelengths. This is because the *SNR* is directly related to the sky transmission, which is determined by the phases of each planet for the imaging and the null baselines. As a result, the reconstructed spectrum of each planet is very noisy in most of the observing wavelength range.

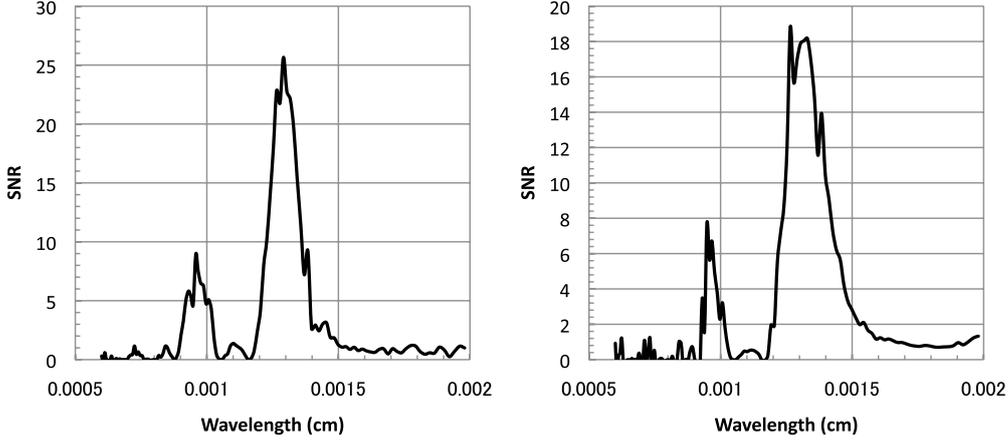
Figure 6. Wavelength dependency of the S/N for the analog Earth (left) and the super-Earth (right).

In order to decrease the wavelength dependency of the *SNR*, we introduce an additional observation. We rotate the array such that $\varphi_{B2}$ is equal to the azimuth angle of the analog Earth from the *x*-axis, $\varphi_{P1}$. The separations projected on to the imaging baseline vector in the fourth measurement are 0.68 AU for the analog Earth and 1.92 AU for the super-Earth. Here, thanks to the fourth measurement, the number of the combinations of four measurements increases to four. Figure 7 shows the *SNR*s of each planet for the combinations of four measurements. Figure 8 shows the averaged *SNR* of each planet, which is estimated according to Equation (15). The averaged *SNR* of each planet improves over the whole observing wavelengths except for around 11.5μm. The dip in *SNR* around 11.5μm results not from the intrinsic behavior of this method but is an artifact of the observing parameters. Figure 9 shows the *SNR*-weighted spectra of the two target planets. These are much closer to the model spectra, compared with those derived from the observations on only three observations on only three observations, but the spectrum of each planet is still noisy at 11.5μm. The observing time in each baseline is 10 days and the total integration time is 40 days.

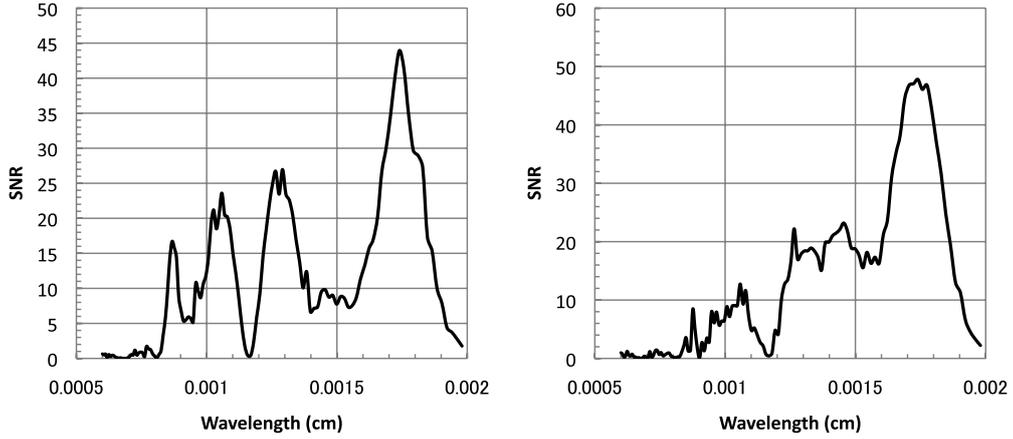
Figure 7. SNRs of the analog Earth (left) and the super-Earth (right) in combinations of four measurements. The black solid, black dashed, gray solid, and black dotted lines are, respectively, S/Ns in the 1–2–3, 1–3–4, 1–2–4, and 2–3–4 measurements, where i−j−k represents the combination of the *i*th, *j*th, and *k*th measurements.

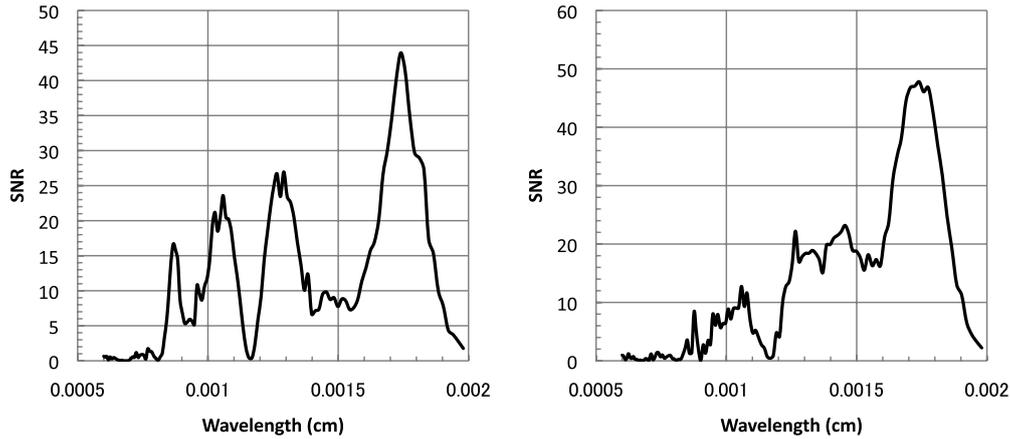
Figure 8. Average of the three SNRs for the analog Earth (left) and the super-Earth (right).

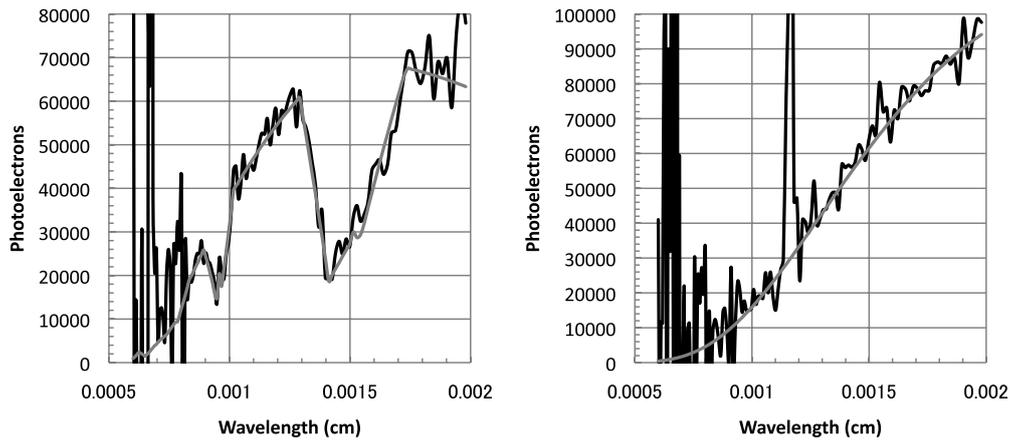
Figure 9. Reconstructed spectra (black solid line) of the Earth-like planet (left) and the Jupiter-like planet (right). The gray solid lines represent the model spectrum.

## 4. CHARACTERIZATION

In the previous section, the positions of analog Earth and super-Earth were extracted and those spectra were also reconstructed based on the proposed method. In this section, we evaluate the reconstructed spectra and examine whether this method allows us to characterize the planets.

### 4.1 Biomarkers

There are various absorption lines in the mid-infrared: $CH_4$ at 7.5μm, $O_3$ at 9.6μm, $H_2O$ around 6μm and beyond 16μm, $CH_4$ at 7.5μm, and several nitrogen compounds, $N_2O$ at 7.8, 8.5, and 17μm, $NH_3$ at 9.6 – 11μm, NO at 5.4μm, and $NO_2$ at 6.2μm. Detection of these absorption lines allows us to examine the existence of biological activity on planets (e.g., Selsis et al. 2005). The $O_3$ composition is the most important for an indication of biological activity because oxygenic photosynthesis in the biosphere, the dominant metabolism of cyanobacterias and plants, builds up the current $O_2$-rich atmosphere on the Earth. Here, an $O_2$ molecule is photochemically decomposed into two O atoms by UV radiation. A chemical reaction among the O atoms, the $O_2$ molecules, and any compound, called a three-body reaction, results in buildup of an $O_3$ atmosphere. $O_3$ is a tracer of $O_2$, but the amount of $O_3$ depends only weakly on the total amount of $O_2$ (Leger et al. 1993).

Selsis et al. (2002) investigated a possibility of abiotic synthesis of $O_2$ and $O_3$ and found that $O_2$ and $O_3$ can be more easily formed instead of the fast recombination of $CO_2$ as the abundance of $CO_2$ increases in the atmosphere. According to their simulation on an accumulation of $O_2$ and $O_3$ through the abiotic processes in the atmosphere on exoplanets, a

$CO_2$ absorption band appears at 9.4 and 10.5μm and also masks the ozone absorption line at 9.6μm when the $CO_2$ pressure is more than 50 mbar. Therefore, a much higher spectral resolution (R = 100) can separate the $O_3$ band at 9.6μm from the $CO_2$ band at 9.4 and avoid a false positive detection. The other absorption lines of CH4, NH3, and NO2 are still questionable biological signatures because the possibility of abiotic synthesis of these compounds has not been fully studied. Indeed, an atmospheric level of CH4 equal to that of the Earth can also be supplied through an impact of cometary ice, which contains 0.5%–1% of CH4 (see Selsis et al. 2005).

### 4.2 SNR

Based on this discussion, we focus on $H_2O$, $O_3$, and $CO_2$ bands and examine how much SNR is required for detection of these bands. The depths of the absorption features are characterized by the atmospheric profiles of abundance and temperature. For example, the depth of the ozone band strongly depends on a temperature contrast between the surface and the stratosphere. The depth of each absorption feature in the current Earth's atmosphere is approximately 0.8 and 0.2 for $H_2O$ around 6μm and beyond 16μm, 0.5 for $O_3$ at 9.6μm, and 0.6 for $CO_2$ at 15 μm. Therefore, the discrepancy between the reconstructed and the model spectra should be less than ±10% for detection of these strong absorption features.

Figure 10 shows the ratio of the reconstructed to the model spectra for each planet as a function of SNR. The discrepancy between the reconstructed and the model spectra is very large for SNR of less than 5. In other words, the reconstructed spectra are very noisy at the specific wavelengths, where SNR is less than 5. On the other hand, the ratio of the reconstructed to the model spectra is close to 1 as SNR increases. The difference is less than ±10% for most of the data with SNR of more than 5. Conclusively, the SNR required for detection of the strong absorption lines should be more than 5.

We extract only the data with SNR of more than 5 and then reconstruct the spectrum of each planet. Figure 11 shows the spectra of analog Earth and super-Earth with SNR of more than 5. The reconstructed spectrum agrees well with the model one for each planet. $H_2O$, $O_3$, and $CO_2$ absorption features can be detected with a spectral resolution of 100. This method also allows us to resolve the $O_3$ at 9.6 μm and the $CO_2$ bands at 9.4 and 10.5μm appearing in the $CO_2$-rich atmosphere and thus avoid a false positive detection. So it allows us to adequately characterize the analog Earth around the Sun-like star at 10 pc and examine whether biological activity on the planet exists.

### 4.3 Error in the Position of the Planet

The finite imaging baseline length produces the error in the position of each planet. As a result, the error introduces differences between the measured and the model spectra, which can be estimated from Equations (12)–(14). We derive the difference here through numerical simulation. For simplicity, we evaluate the impact of only the error in the position of the planet on the reconstructed spectrum. Figure 12 shows the reconstructed spectrum of the analog Earth in the case that the error in its position is (−0.014 AU, 0.014 AU). The applied value is equal to the maximum ambiguity in the planet's position along each axis determined through the observations on more than three baselines. In this case, the difference between the model and the reconstructed spectra is less than ±10% over the observing wavelengths except for those longer than 19 μm. Therefore, an additional criterion is required for avoiding artificial noise. A study on this criterion will be made in the future.

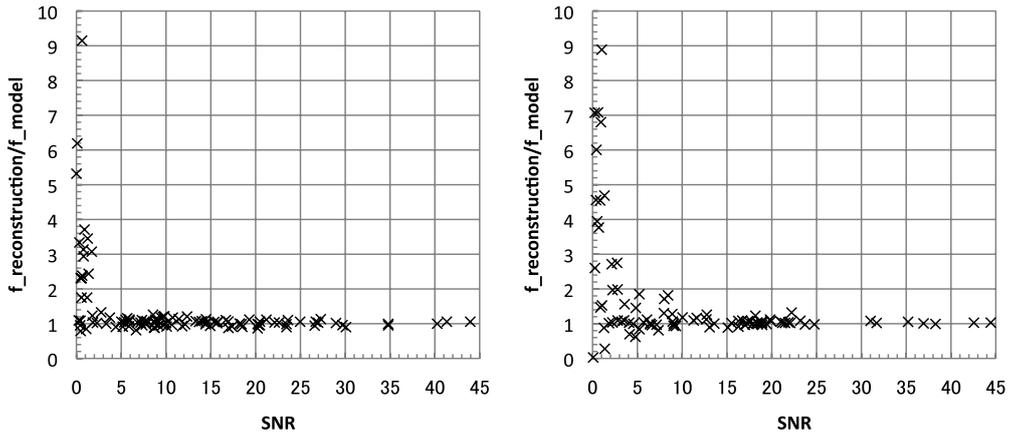

Figure 10. Ratio of the reconstructed spectrum to the model one for each planet as a function of SNR.

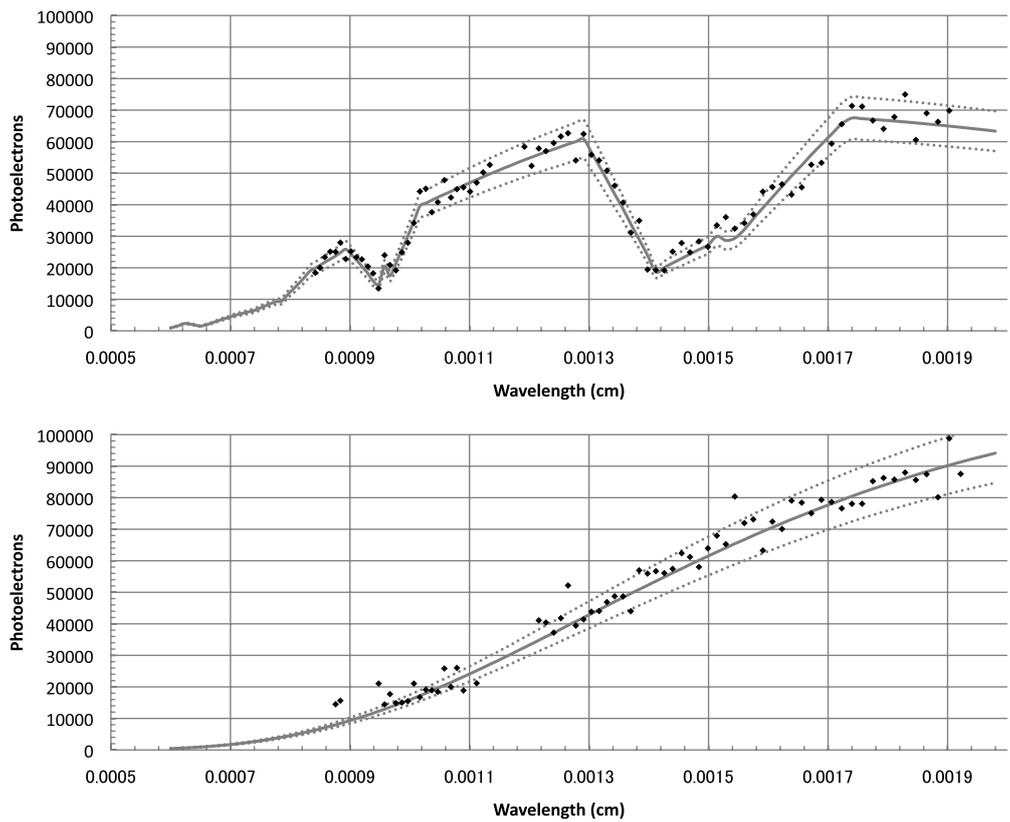

Figure 11. Reconstructed spectra with S/N of more than 5 (dots) for the analog Earth (upper) and the super-Earth (lower). The gray solid and the dotted lines represent the model spectrum and 0.9 or 1.1 times its spectrum, respectively.

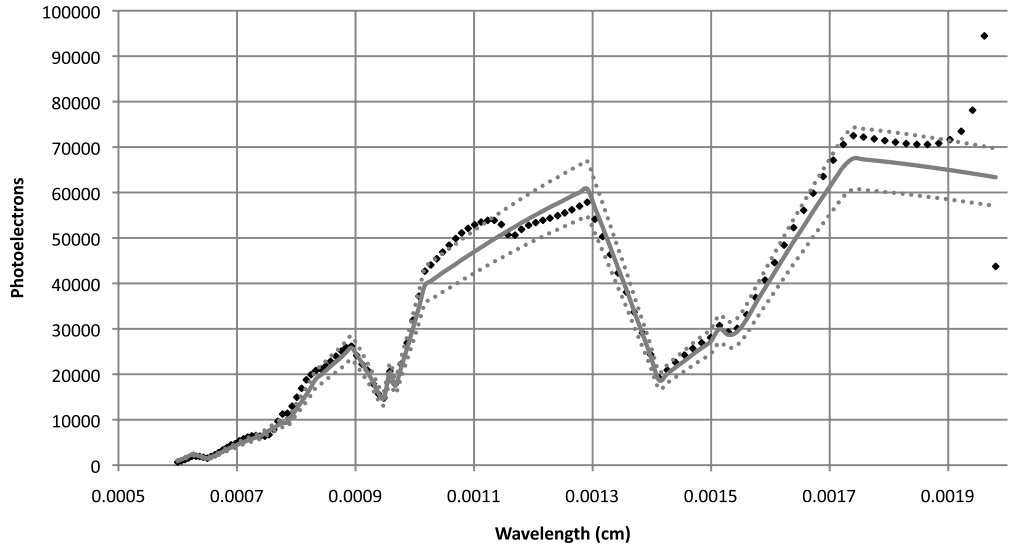

Figure 12. Impact of the error in the position of the analog Earth on its reconstructed spectrum (dots). The gray solid and the dotted lines represent the model spectrum and 0.9 or 1.1 times its spectrum, respectively.

## 5. CONCLUSION

In this paper, we presented a novel concept for direct detection and characterization of exoplanets in the mid-infrared. There are several advantages to this method. We show that the (u,v) plane can be simultaneously and uniformly filled by use of the continuous observing wavelengths instead of continuously rotating the baselines. As a result, this concept allows us to acquire a clean reconstructed image and to perfectly remove speckles from the reconstructed images. Furthermore, observations on only three and n + 1 baselines respectively are sufficient to extract an image of the planetary system and the spectrum of each planet, although the SNR will improve with more baselines. We validated the concept using numerical simulations. Two target planets, an analog Earth and a super- Earth, orbiting a Sun-like star at 10 pc were taken as an example. The position of each planet on the sky and its spectrum were derived through this method. The analog Earth and the super-Earth could be detected from the observations on three baselines and the spectrum of each planet could be successfully obtained with a spectral resolution of 100 from observations on four baselines.

We also discussed characterization of the planets in the mid- infrared and mentioned the possibility of detection of false biological features produced by an abiotic process. Based on current studies, a simultaneous signature of $H_2O$, $O_3$, and $CO_2$ is a potential biological indicator. On the other hand, in a $CO_2$-rich atmosphere, $O_2$ and $O_3$ could be photochemically accumulated through photolysis of $CO_2$. In addition, two $CO_2$ absorption bands appear at 9.4 and 10.5 μm and then mask the $O_3$ absorption feature at 9.6 μm. We found that an SNR of 5 or more is required for secure detection of $H_2O$, $O_3$, and $CO_2$ absorption features. The reconstructed spectrum can be acquired at observing wavelengths from 8 to 19 μm for the observing time of 40 days. Thanks to the high spectral resolution R = 100, this method also resolves the $O_3$ and the $CO_2$ bands produced in the $CO_2$-rich atmosphere and avoids a false positive detection. Thus, this method allows us to fully characterize the planet and examine whether biological activity exists.

We need further analysis of this concept, but already the basic idea is clear: measurements on a few discrete baselines are sufficient to extract images and spectra from multi-planet systems. We also need to include the effect of asymmetries in the exozodiacal distributions: Defre`re et al. (2010) estimated the impact of these asymmetrical disks on the previous method and concluded that this effect could produce artificial planet signals. We reserve these modifications for a future paper.

## ACKNOWLEGEMENT

We very much thank an anonymous referee whose sugges- tions improved this study. In this work, T.M. was supported as a Research Fellow of the Japan Society for the Promotion of Science, and he thanks the Jet Propulsion Laboratory for its hospitality.